\documentclass[twocolumn,superscriptaddress,showpacs]{revtex4-1}
\usepackage{graphicx}
\usepackage{color}
\usepackage{multirow}
\usepackage{rotating}
\usepackage{amsmath}

\begin{document}

\title{
Quantum Monte Carlo study of\\
one-dimensional transition metal organometallic cluster systems\\
and their suitability as spin filters
}

\author{L. Horv\'athov\'a}
\author{R. Derian}
\affiliation{
Institute of Physics, CCMS, Slovak Academy of 
                      Sciences, 84511~Bratislava, Slovakia}
\author{L. Mitas}
\affiliation{
North Carolina State University, Department of Physics, Raleigh, NC 27695} 

\author{I. {\v S}tich}
\email{ivan.stich@savba.sk}
\affiliation{
Institute of Physics, CCMS, Slovak Academy of 
                      Sciences, 84511~Bratislava, Slovakia}

\date{\today}

\begin{abstract}
We present calculations of electronic and magnetic structures of  vanadium-benzene multidecker clusters V$_{n}$Bz$_{n+1}$
 ($n$ = 1 - 3) using advanced quantum Monte Carlo methods. 
These and related systems have been identified as prospective spin filters in spintronic applications, assuming that their ground states are half-metallic ferromagnets. Although we find that magnetic properties
of these multideckers are consistent with ferromagnetic coupling, their electronic structures do not appear to be half-metallic as previously assumed. In fact, they are ferromagnetic insulators with large and broadly similar $\uparrow$-/$\downarrow$-spin gaps. This makes the potential of these and related materials as spin filtering devices very limited, unless they are further modified or functionalized.                                         
\end{abstract}


\maketitle

\section{\label{sec:intro}Introduction}

\begin{figure}[t!]
\centering
\includegraphics[width=0.55\columnwidth,clip,angle=0]{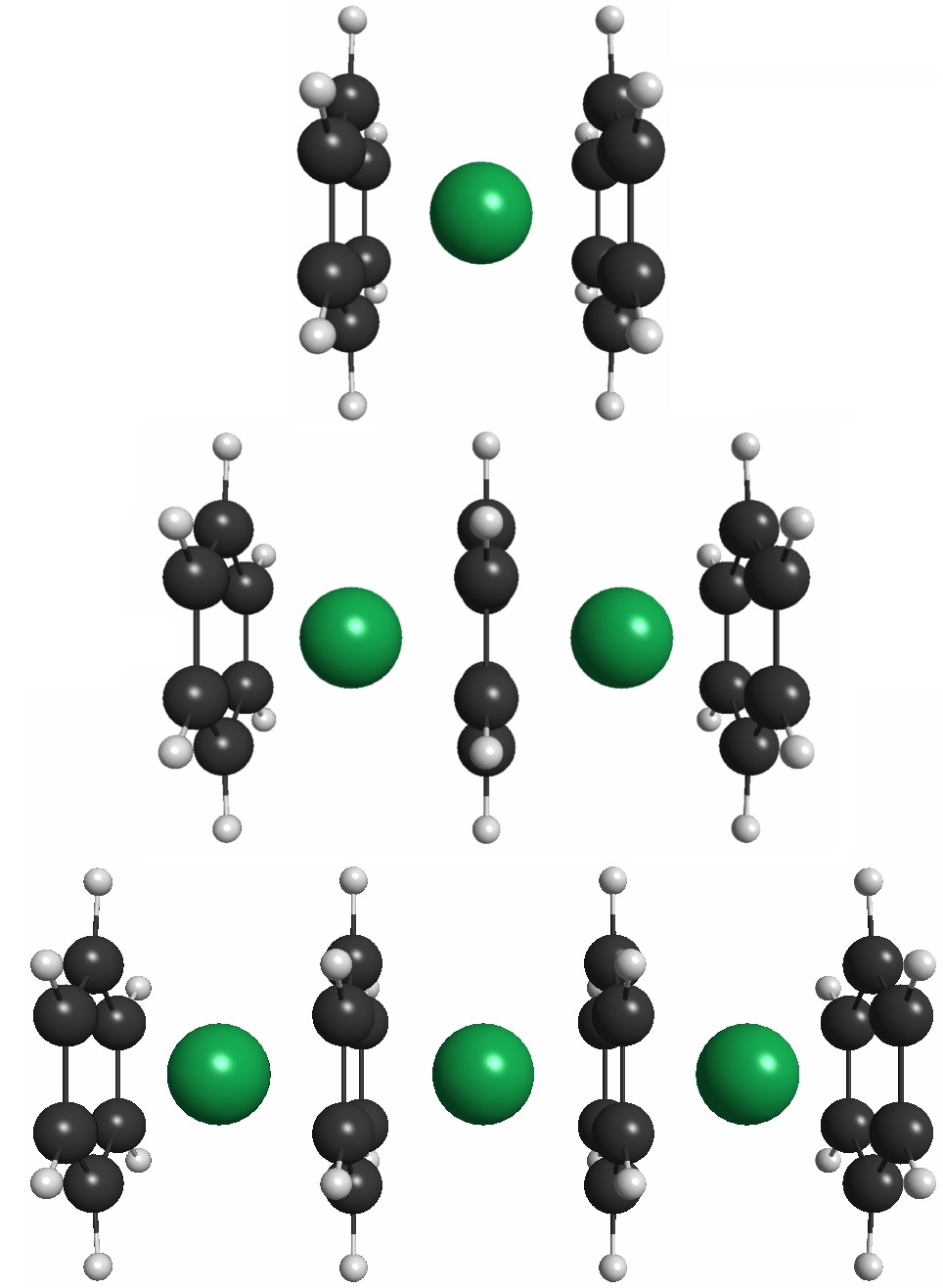}\\
\vspace{0.5cm}
\includegraphics[width=1.0\columnwidth,clip,angle=0]{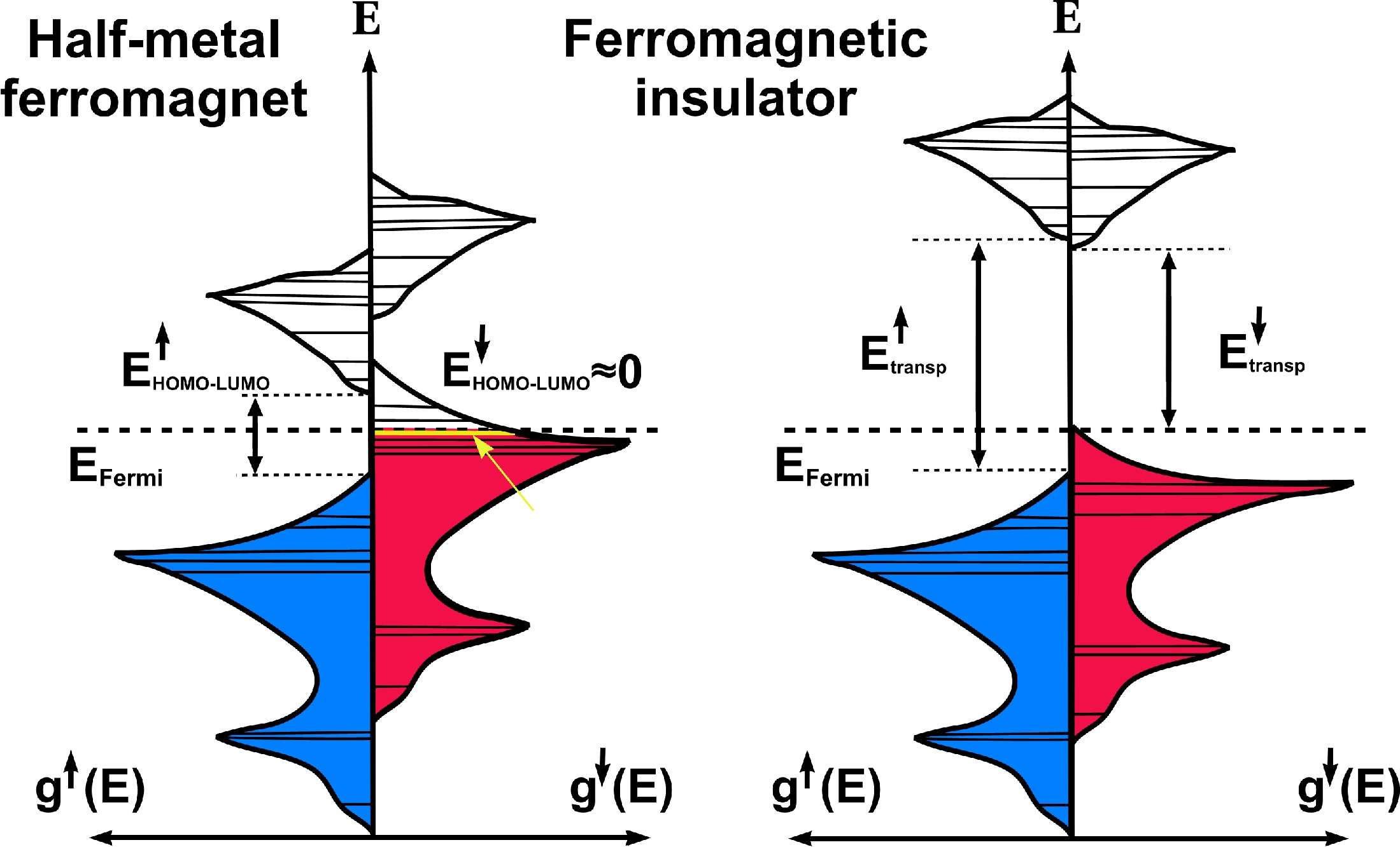}
\caption{ 
Sandwich-like organometallic structures of VBz complexes with transition metal atoms separated by benzene molecules (top). Bottom: Schematic illustration of a half-metal ferromagnet featuring a single spin-channel at the Fermi energy (yellow color and arrow) in the minority spin population and of a ferromagnetic insulator featuring much larger transport gaps $E_{transp}^{{\uparrow}/{\downarrow}}$. For a more realistic DFT $g^{\uparrow}(E)$/$g^{\downarrow}(E)$, see Ref.~\cite{blugel}.
}
\label{fig:scheme_sandwich}
\end{figure}

As predicted long time ago \cite{mott_36}, electrons with up- and down-spins can exhibit different resistivities in different spin channels,  $R^{\uparrow} \ne R^{\downarrow}$. This can happen in special types of materials with particular electronic structure features that favor transport of one spin orientation over the other.
Over the past decade, this property has been exploited in spintronics~\cite{fabian_04} combined with molecular electronics by employing both nonmagnetic~\cite{ratner_10} and magnetic bridges~\cite{sanvito_11}. 
In particular,
half-metallic (HM) ferromagnet (FM) with high Curie temperature and only one electronic spin channel at the Fermi energy~\cite{deGroot_83} would be an ideal realization of a molecular spin filter. Such behavior have recently been predicted for transition metal organometallic sandwich structures~\cite{jellinek_05,maslyuk,blugel,feng_08,sanvito_08,yang_09,huang_13,nakajima_12,kashyap_13} (Fig.~\ref{fig:scheme_sandwich}) with promising perspectives for applications in spintronics~\cite{sanvito_11} and quantum computing~\cite{loss_01}. These conclusions have been based mainly on computational studies since such behavior of metal-organometallic molecule junctions has not been experimentally demonstrated yet. 
In order to shed a fresh light on this problem, we employ advanced quantum Monte Carlo (QMC) methods~\cite{QMC_method}
to study the transition metal-benzene multidecker molecules V$_{n}$Bz$_{n+1}$, n = 1 - 3,  as one of the most prominent
examples of prospective spin filtering systems.  Our methods raise the accuracy and predictive power of the calculations
by about an order of magnitude with regard to previous studies and enable us to study subtle effects of spin flips and corresponding energy shifts with much higher degree of confidence.  Based on our results, we conclude that these systems are ferromagnetic insulators with large and broadly similar $\uparrow$-/$\downarrow$-spin gaps, see Figure~\ref{fig:scheme_sandwich}. We do not find 
half-metallic ferromagnets as the previous calculations, implying thus a
limited potential of these materials for spintronic applications unless they are further modified or functionalized.                                                 

\vspace{0.5cm}
In addition to the anticipated applications in spintronics and quantum computing, transition metal-benzene (TMBz) multidecker molecules of the TM$_{n}$Bz$_{n+1}$ type have attracted attention also because of their potential for other applications in catalysis, polymers, molecular recognition~\cite{sand_clust_review}, and high-density storage. 
Mass spectra experiments~\cite{kurikawa} on positively charged TMBz$^{+}$ complexes and theoretical studies~\cite{pandey_2} on both charged and neutral clusters suggest that only systems with the early TMs adopt the multidecker molecular structure, whereas the late TMs form rice-ball structures where TM atoms are fully covered with Bz molecules. One of the most studied early transition metal-benzene multidecker molecules, V$_{n}$Bz$_{n+1}$~\cite{jellinek_05,xiang,maslyuk,blugel,nakajima_12}, was expected to exhibit the desired HM/FM properties for sandwiches of medium length (n $\approx$ 3) with minority ($\downarrow$) spin-gaps featuring metallic behavior and majority ($\uparrow$) spin-gaps featuring semiconducting behavior. Experimentally the multidecker VBz molecules were synthesized up to 
$n$ = 6 in a reaction of laser-vaporized V atoms with benzene molecules~\cite{dougherty_1,dougherty_2} and, hence, cover also the anticipated HM/FM range. The FM nature of the V$_{n}$Bz$_{n+1}$ sandwiches was predicted both experimentally in Stern-Gerlach (SG) magnetic deflection experiments~\cite{stern-gerlach_2} and theoretically in density functional theory (DFT) modeling~\cite{jellinek_05,xiang,maslyuk,blugel,nakajima_12}. However, to the best of our knowledge, the HM behavior was only predicted theoretically in DFT calculations~\cite{jellinek_05,xiang,maslyuk,blugel,nakajima_12} and experiments that would measure spin transport through metal-V$_{n}$Bz$_{n+1}$-metal junction are still missing.

The identification of V$_{n}$Bz$_{n+1}$ sandwiches as half-metal ferromagnets has several weak points. The SG experiment revealed that the magnetic moment increases monotonically with the cluster size up to $n$ = 4~\cite{stern-gerlach_2}, supporting thus the FM ordering. However, the experimental error bars in the region of interest (n $\approx$ 3) of $\approx\pm$1 preclude an unambiguous interpretation of the magnetic state of the molecule. In addition to ferromagnetic states, there might  be also anti-ferromagnetic (AFM) states with similar energies. Such states have been studied using DFT techniques that predict V$_{n}$Bz$_{n+1}$ multidecker molecules to be either FMs~\cite{sanvito_08} or FMs and AFMs, depending on $n$~\cite{blugel}, with energy differences of the order of few meV. 
The competition between the direct exchange between the V atoms favoring ferromagnetism and super-exchange via the benzene ring favoring antiferromagnetism was identified as the reason for these small energy differences~\cite{blugel}.
However, limited accuracy and reliability of DFT techniques for magnetic states of TM-based organometallics have been known for some time. For example, comparisons with high-accuracy QMC calculations showed qualitatively incorrect, strongly exchange-correlation (xc) model-dependent DFT calculations of magnetic states for TMBz half-sandwich structures ~\cite{PRL_orgmet,JCTC_orgmet}. Hence, the ability of DFT techniques to correctly predict magnetic structure of full-sandwich VBz multidecker structures appears to be far from clear. The other shortcoming of the DFT methods is their reliance on the highest occupied-lowest unoccupied molecular orbital (HOMO-LUMO) model of the spin gaps, which are of central importance for identification of the HM nature of the V$_{n}$Bz$_{n+1}$ systems.  In fact, there are additional complications in DFT estimations of the electronic gaps besides 
the poor approximation of the optical gaps by the DFT HOMO-LUMO differences.
 Due to the presence of strongly bonded excitons,
there may be a huge difference between optical and transport gaps that are of the key importance for a molecule to function as a spin filter. 
Hence, in order to uncover the true nature of the magnetic and electronic structures in the V$_{n}$Bz$_{n+1}$ multidecker molecules we have carried out a very accurate investigations of the gaps  for $n$ = 1 - 3. This covers the range of the anticipated half-metal ferromagnet behavior and sorts out the experimental and theoretical uncertainties mentioned above. Our calculations, with accuracy higher by an order of magnitude when compared to previous DFT-based studies, predicts V$_{n}$Bz$_{n+1}$ clusters to be \textit{ferromagnetic insulators} rather than half metals.

\section{\label{sec:methods}Methods}
\hspace{-4mm}The details of our calculations are in Supplementary Information (SI). Briefly, we use a four-level refinement strategy that enabled us to eliminate all the systematic biases, except for the fixed-node approximation~\cite{QMC_method}: 1) initial geometries were obtained from DFT optimization, 2) the single-determinant trial wave function was constructed from spin-unrestricted DFT single-particle states using Becke-Perdew-Wang~91~\cite{becke,perdew-wang} xc-functional, 3) the trial wave function was optimized using VMC (variational Monte Carlo) techniques, 4) final energies were computed from the fixed-node DMC (diffusion Monte Carlo) simulations. Spin-unrestricted results exhibited only minor spin-contamination, see Supplementary Tables I-III. We also note that QMC attempts for structural reoptimization left the DFT-optimized structures unchanged, c.f. Supplementary Figure 1. These tests justify use of spin-unrestricted DFT wave functions and DFT-optimized structures in our QMC calculations. For static DFT modeling we used the \texttt{GAMESS} suite of codes~\cite{gamess1}, while all VMC and DMC calculations used the \texttt{QWalk} code~\cite{qwalk}. Based on our previous experience with VBz half-sandwiches~\cite{PRL_orgmet,JCTC_orgmet}, we use the gradient-corrected (GGA) BPW91~\cite{becke,perdew-wang} 
DFT xc-functional, for alternative choices see SI. One of the few direct comparisons with experiments available for neutral V$_{n}$Bz$_{n+1}$ multidecker molecules are vertical ionization potentials (IP)~\cite{kurikawa}. IPs are very accurately experimentally measurable and we used IPs to benchmark the methods we use in our study. We compute the following IPs for V$_{n}$Bz$_{n+1}$: 5.71(4), 4.70(7), 4.06(9) eV, for n = 1, 2, and 3, compared to the experimental values of 5.76(3), 4.70(4), 4.14(5) eV~\cite{kurikawa}. The importance of IPs is also in the fact that, in the spin-polarized version, they enter calculation of the spin transport gaps, see Section~\ref{sec:results}. Further details are presented in Supplementary Table VIII and Supplementary Figure 2.

\section{\label{sec:results}Results and discussion}

\begin{table}
\centering
\caption{V$_{n}$Bz$_{n+1}$ $\to$ nV + (n+1)Bz adiabatic fragmentation energies per V atom (in eV) at different levels of treatment.}
\scriptsize
\begin{tabular}{l|c|c|c}
\hline
\hline
method            & VBz$_{2}$   & V$_{2}$Bz$_{3}$ & V$_{3}$Bz$_{4}$\\
\hline
DFT         & 5.24    &4.78     & 4.60 \\ 
DMC         & 4.28(4) &3.80(6)  & 3.61(7) \\
\hline
\hline
\end{tabular}
\label{tab:vbz_dissoc_energy}
\end{table}

First, we analyze chemical bonding in the V$_{n}$Bz$_{n+1}$ sandwich systems. Adiabatic fragmentation energies in the V$_{n}$Bz$_{n+1}$ $\to$ nV + (n+1)Bz channel are shown in Table~\ref{tab:vbz_dissoc_energy}; total energies of all V$_{n}$Bz$_{n+1}$ complexes and their fragments are summarized in Supplementary Tables I-VI and energies of other dissociation channels can be found in Supplementary Table VII. 
The impact of electron correlation effects captured in QMC treatment is clearly visible.  
The -Bz$\cdots$V$\cdots$Bz- units are strongly bonded with binding energies $>$3.5 eV or 1.75eV per V$\cdots$Bz bond.
These per V atom energies represent averaged values. More detailed analysis exposes strong asymmetries in the V$\cdots$Bz bonding between terminal and inner molecular bonds. For example, DMC adiabatic dissociation energy of the terminal Bz-V$\cdots$Bz-(VBz)$_{2}$ bond is $\approx$2.5eV, whereas it is only $\approx$1eV for one of the inner Bz-V-Bz$\cdots$V$\cdots$Bz-V-Bz bonds, see Supplementary Table VII.
Given the fact that experimentally the dissociation energies of the neutrals are derived from dissociation energies measured for V$_{n}$Bz$_{n+1}^{+}$ cations, corrected for kinetic shifts, using vertical metal atom and neutral complex ionization energies~\cite{kurikawa,pandey_2}, the QMC-calculated adiabatic dissociation energies are expected to be more accurate than the indirectly determined experimental values~\cite{PRL_orgmet,JCTC_orgmet}.

\begin{figure}[b!]
\centering
\includegraphics[width=1.0\columnwidth,clip,angle=0]{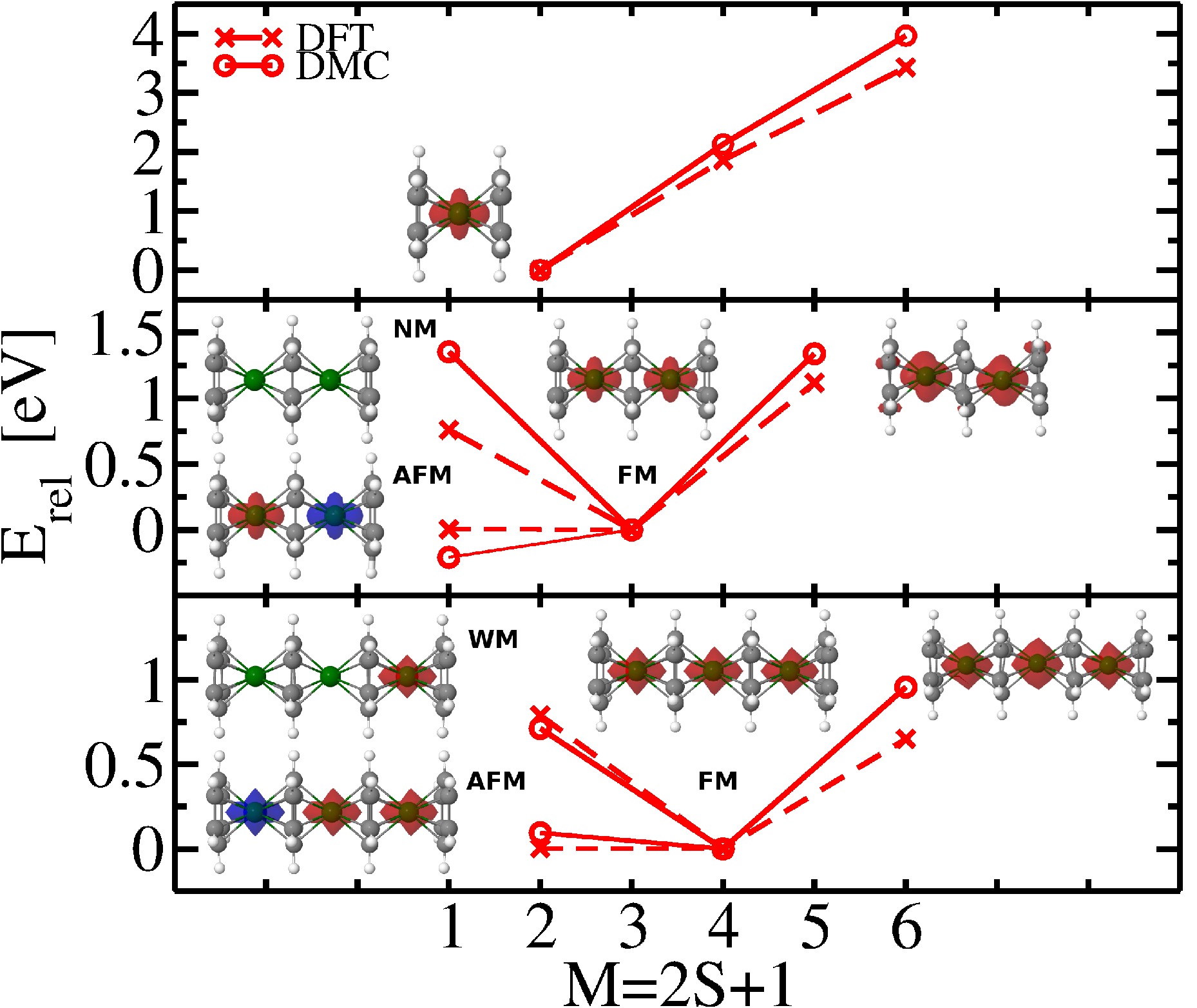}
\caption{ 
Relative energies of V$_{n}$Bz$_{n+1}$, $n$ = 1 - 3 as a function of spin multiplicity. The insets show spin densities, for more details see the text.   
The error bars in DMC results are comparable to the size of the symbols.
}
\label{fig:energies_VBz}
\end{figure}
 
Next, we determine magnetic ordering by calculating total energies of the V$_{n}$Bz$_{n+1}$ multidecker molecules for $n$ = 1 - 3 in a variety of different spin states, including nonmagnetic (NM), weakly magnetic with only one unpaired electron (WM), 
FM, and AFM states. 
In addition to these \lq\lq natural\rq\rq~spin states we also consider high-spin ferromagnetic states with unusual spin multiplicities, such as, for example, M = 6 in V$_{3}$Bz$_{4}$, that would correspond to five unpaired electrons distributed over three vanadium sites. 
SG experiments~\cite{stern-gerlach_2} suggest a linear increase of the magnetic moment $\mu_{z}$ with $n$, corresponding to ferromagnetic coupling with each vanadium atom contributing spin $\frac{1}{2}$.
Our calculated results shown in Figure~\ref{fig:energies_VBz} indicate that the FM and AFM states are very close in energy, degenerate within our error bars.
The small energy differences are most likely due to the competition between direct and super-exchange between the magnetic centers~\cite{blugel}.
This small energy scale prevents us from determining accurate exchange energy~\cite{reboredo_14} for the vanadium centers and Curie temperatures which could be measured experimentally, for further discussion, see SI.
Both non-magnetic and high-spin states are significantly, $>$0.5eV, higher in energy than the FM/AFM states, more details can be found in Supplementary Tables I-III. 
FM, AFM, and WM states all feature spins in $d_{z^{2}}$ orbitals oriented along the molecular axis, see Figure~\ref{fig:energies_VBz}. 
While our accuracy is not sufficient to discriminate clearly between FM and AFM states,  the results are consistent with SG experiments~\cite{stern-gerlach_2} that suggest the ground states of multiplicity of M =  2, 3, and 4 for $n$ = 1, 2, and 3, hence,
 \textit{ferromagnetically} coupled magnetic moments of 1 $\mu_{B}$ per the V atom.

\begin{figure}[t!]
\centering
\includegraphics[width=1.0\columnwidth,clip,angle=0]{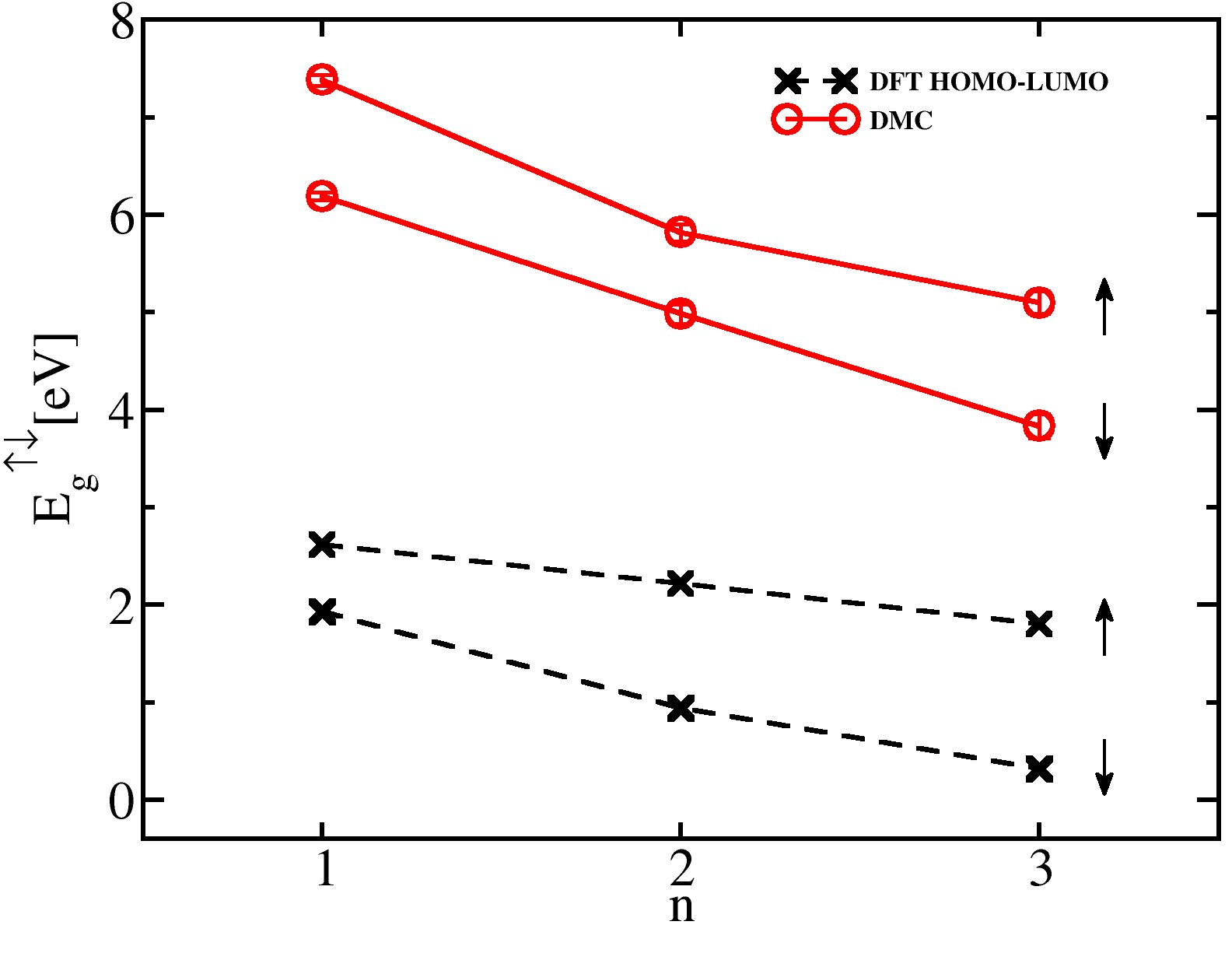}
\caption{ 
Comparison of calculated DFT HOMO-LUMO spin-gaps and DMC transport spin-gaps calculated from formula (\ref{eq:spin_gaps}) for V$_{n}$Bz$_{n+1}$, 
$n$ = 1 - 3, in the ferromagnetic state.
}
\label{fig:spingaps_VBz}
\end{figure}

Spin filtering function of the ferromagnetic V$_{n}$Bz$_{n+1}$ molecules requires that they exhibit also the half-metallic behavior. For this purpose, we have evaluated the fundamental spin-gaps of V$_{n}$Bz$_{n+1}$ in the range $n$ = 1 - 3. For comparison, we present also the DFT HOMO-LUMO gaps that are routinely used to estimate quantum transport~\cite{jellinek_05,maslyuk,blugel,feng_08,sanvito_08,yang_09,huang_13,nakajima_12,kashyap_13} and that are straightforward to calculate. Since the HOMO-LUMO gap is a single-particle quantity, in the many-body QMC method we calculate instead the vertical transport spin-gaps $E_{transp}^{{\uparrow}/{\downarrow}}$, Figure~\ref{fig:scheme_sandwich}, as differences between the energies with an added or subtracted spin $\uparrow$-/$\downarrow$-electron ~\cite{QMC_method} as follows 
\begin{equation}\label{eq:spin_gaps}
E_{g}^{\uparrow,\downarrow} = (E_{N+1}^{\uparrow,\downarrow} - E_{N}) -  (E_{N} - E_{N-1}^{\uparrow,\downarrow})~~~.
\end{equation}
Here $E_{g}^{\uparrow,\downarrow}$ is the sum of the spin-$\uparrow$/$\downarrow$ vertical ionization potential and electron affinity, which can be fairly easily and very accurately calculated by QMC methods. The calculated DFT HOMO-LUMO and DMC vertical transport spin-gaps for the respective ground-state multiplets are in Figure~\ref{fig:spingaps_VBz}; more details, including discussion of adiabatic transport spin-gaps and results obtained with other DFT xc-functionals, can be found in Supplementary Table IX and Supplementary Figure 3. The DFT HOMO-LUMO spin-gaps indeed reproduce the previously reported HM nature around $n$ = 3~\cite{jellinek_05,maslyuk,blugel,huang_13,nakajima_12} with minority ($\downarrow$) spin-gaps featuring metallic behavior and majority ($\uparrow$) spin-gaps featuring semiconducting behavior. Qualitatively similar results are obtained also with other DFT xc-functionals, see SI. However, the DMC transport spin-gaps feature a completely different behavior, with \textit{insulating} spin-gaps for both spin types, see Figure~\ref{fig:spingaps_VBz}, with large ($>$4 eV) and broadly similar $\uparrow$-/$\downarrow$-spin gaps. As indicated in the introduction, the qualitatively incorrect conclusion drawn from the HOMO-LUMO spin gaps is due to a combination of two effects: a) underestimation of the optical gap by the DFT HOMO-LUMO gap, and b) neglect of the excitonic binding energy, that is often large in organic materials. Hence, we conclude that the family of V$_{n}$Bz$_{n+1}$ multidecker molecules are \textit{ferromagnetic insulators} and \textit{not} half-metal ferromagnets, as heretofore assumed. 

\section{\label{sec:conclusions}Conclusions}

In summary, we report the most accurate study to date of magnetic ordering and electronic structure of the V$_{n}$Bz$_{n+1}$ multidecker clusters. We use a combination of DFT and QMC methods
with a boost in accuracy of predictions by at least an order of magnitude when compared
with previous studies.
Among the many potential applications of these materials we have focused on properties relevant to spintronics and quantum computing. Of the key importance for such applications is the half-metal ferromagnetic nature of the molecule.  Contrary to the large number of previous studies, 
we conclude that this class of materials are \textit{ferromagnetic insulators} with large and broadly similar $\uparrow$-/$\downarrow$-spin gaps, as opposed to \textit{half-metal ferromagnets}, as hitherto assumed~\cite{jellinek_05,maslyuk,blugel,huang_13,nakajima_12}. One could argue that the HM/FM limit could be achieved for longer molecules, around $n \approx$ 7. However, V$_{n}$Bz$_{n+1}$ molecules have experimentally been synthesized only up to $n$ = 6~\cite{dougherty_1,dougherty_2}. Moreover, V$_{n}$Bz$_{n+1}$ molecules for $n \ge$ 4 have been reported to assume chiral structure~\cite{jellinek_05} that leads to opening of the electronic gap. Since all the predictions of HM/FM behavior also for the related 1-dimensional multidecker TM organometallics~\cite{feng_08,sanvito_08,yang_09,kashyap_13} are based on essentially the same type of analysis, they are expected to suffer from the same shortcomings. We conclude that the potential of these systems to be employed as spin filters is limited although they could be useful for other applications \cite{sand_clust_review} and/or need to be further modified in order to achieve the required electronic structure properties.

\begin{acknowledgments}
This research was supported in part by APVV-0207-11, LPP-0392-09, and VEGA (2/0007/12) projects, and via CE SAS QUTE. 
We also gratefully acknowledge the Computing Centre of the Slovak Academy of Sciences and use of the supercomputing infrastructure funded by the ERDF, projects ITMS 26230120002 and 26210120002. We also acknowledge support by ARO W911NF-04-D-0003-0012 and NSF OCI-0904794 grants as well as XSEDE allocations at TACC.
\end{acknowledgments}

\bibliography{VBz_fullsandw}

\end{document}